\documentclass[titlepage, aps,prb,twocolumn]{revtex4}

\usepackage{graphicx}
\usepackage{amsmath}
\usepackage{xcolor}
\usepackage{physics}

\usepackage[pdftex,colorlinks=true]{hyperref}
\hypersetup{
linkcolor=blue,
citecolor = blue,
urlcolor=blue
}

\begin{document}

 
\setlength{\pdfpagewidth}{8.5in}
\setlength{\pdfpageheight}{11in}

\title{Theory of the low- and high-field superconducting phases of UTe$_2$}

\author{Josephine J. Yu$^{1}$, Yue Yu$^{2}$, Daniel F. Agterberg$^{2}$,  S. Raghu$^{3,4}$}

\affiliation{$^{1}$ Department of Applied Physics, Stanford University, Stanford, CA 94305, USA \looseness=-1}
\affiliation{$^{2}$ Department of  Physics,University of Wisconsin-Milwaukee, Milwaukee, WI 53211, USA \looseness=-1}
\affiliation{$^{3}$ Stanford Institute for Theoretical Physics, Stanford CA, 94305 USA \looseness=-1}
\affiliation{$^{4}$ Department of  Physics, Stanford University, Stanford, CA 94305, USA \looseness=-1}

\date{\today}

\begin{abstract}
Recent nuclear magnetic resonance (NMR) and calorimetric experiments have observed that UTe$_2$ exhibits a transition between two distinct superconducting phases as a function of magnetic field strength for a field applied along the crystalline $b$-axis. To determine the nature of these phases, we employ a microscopic two-band minimal Hamiltonian with the essential crystal symmetries and structural details. We also adopt anisotropic ferromagnetic exchange terms. We study the resulting pairing symmetries and properties of these low- and high-field phases in mean field theory. 
\end{abstract}
\maketitle

\section{Introduction}


The material UTe$_2$ \cite{Ran2019} has been the subject of extensive recent investigation due to its multifarious manifestations of exotic superconductivity. The upper critical field along the crystalline $b$ axis, $H_c = 40 \text{ T}$\cite{Ran2019, Helm2022}, is strikingly large in light of the critical temperature $T_c = 1.6 \text{ K}$, and is indicative of an odd parity superconducting ground state. UTe$_2$ belongs to a larger family of uranium-based candidate unconventional superconductors, the rest of which exhibit ferromagnetism coexisting with superconductivity \cite{Saxena2000, Aoki2001, Huy2007, Aoki2012}.  In contrast, UTe$_2$ lacks magnetic order \cite{Ran2019}.  Thus, UTe$_2$ offers the opportunity to probe unconventional superconductivity in a family of materials without the confounding effects of magnetism. Furthermore, UTe$_2$ is also believed to host exotic phenomena such as reentrant superconductivity\cite{Ran2019, Knebel2019}, broken time-reversal symmetry \cite{Jiao2020, Hayes2021},  and pair density wave (PDW) order \cite{Gu2022}, thus positioning it as a paradigmatic unconventional superconductor. 
 
Like the other uranium-based superconductors, UTe$_2$ exhibits reentrant superconductivity as 
a function of the strength of the magnetic field applied along the crystalline $b$ axis. While this phenomenon was initially attributed to fluctuations near a ferromagnetic quantum critical point, a few studies\cite{Yu2022,Ishizuka2019} offered an alternative explanation: distinct superconducting phases at low and high magnetic field strengths. 

Recent experiments \cite{Kinjo2022, Rosuel2022} have confirmed the existence of  two distinct superconducting phases in UTe$_2$, distinguished primarily by their responses to the orientation of the applied magnetic field. The state at low magnetic field strengths has little sensitivity to the direction of the field. In contrast, the superconducting state at high magnetic field strengths is easily suppressed by tilting the field away from the $b$ axis in either the $a$ or $c$ directions\cite{Kinjo2022, Rosuel2022, Knebel2019}. Thus, a fundamental question regarding the unconventional superconductivity in UTe$_2$ is: What are the pairing symmetries associated with these superconducting phases? 


In this work, we 
provide  concrete predictions for the pairing symmetries of the low- and high-field superconducting phases.
As we describe in Sec. \ref{sec:pairing}, we do this in two ways. In both, we use a minimal Hamiltonian with the essential symmetries and structural details, including spin-orbit coupling. First, we calculate the pair field susceptibility $\chi$, which is defined in Sec. \ref{sec:chi}. This reveals the dominant superconducting tendencies of the normal state as determined by the kinetic energy. Second, we introduce local, anisotropic ferromagnetic pairing interactions and solve the self-consistent mean-field gap equation with these interactions (Sec. \ref{sec:Delta}). In both approaches, we find that the low- and high-field superconducting states are odd-parity states, but with the spin pointing in different primary directions. At small magnetic field strengths, the pairing state has spin predominantly in the $ac$ plane, but at high enough magnetic field strengths, this spin aligns with the $b$ axis. The change in pairing symmetry also has consequences for the physical properties of the state. In Sec. \ref{sec:properties}, we infer the nodal structure of the gap from our results and discuss the properties of the phases.

\section{Model}
In this section, we describe the tight-binding Hamiltonian used throughout this work and the symmetry classifications of the allowed pairing states. The main assumption in our work is that the fermions relevant for superconductivity reside on uranium atoms, which is reflected in the pairing symmetry classifications and the minimal Hamiltonian. 

Theoretical predictions for the density of states in UTe$_2$ find that uranium $5f$ orbitals contribute the largest density of states at the Fermi energy \cite{Fujimori2021}, so the uranium electrons are likely the driver of superconductivity in the system. The rung (sublattice) structure is also believed to play a crucial role in determining the electronic and superconducting properties of the material\cite{Xu2019, Shishidou2021,Hazra2022}. Thus, we anticipate that the superconducting properties of UTe$_2$, including the symmetries of the pairing states, should be qualitatively well-captured by a minimal model with orthorhombic symmetry and sublattice structure. 

Though we focus here on UTe$_2$, the structural motif mentioned here is present throughout many other candidate unconventional superconductors; prior work\cite{Hazra2022} has studied the effects of this structure using a complementary approach, modeling the local physics using a Hund's-Kondo model.

\subsection{Symmetries \label{sec:model}}

We first describe the symmetries of the crystal and classify the possible pairing states. In UTe$_2$, the pairs of uranium atoms form ``rungs" of a ladder, oriented in the $\hat{z}$ (crystalline $c$) direction, which build up a body-centered orthorhombic crystal. The pairing symmetry classifications are determined by the usual spin and momentum symmetries, together with the uranium site symmetry. Note that the uranium site symmetry $C_{2v}$ has just one spin-representation, so all local Kramer's pairs (time-reversal symmetry related states) must have the same symmetry. The following symmetry classifications are thus general for any number of local orbitals, though we describe the scenario when there is a single local orbital per uranium site and use the terms orbital and sublattice interchangeably. 

In the absence of a magnetic field ($B=0$), the orthorhombic symmetry group ($D_{2h}$) respects inversion $I$ and mirror plane $M_x$, $M_y$, $M_z$ symmetries, which can then be used to classify the possible pairing states. The degrees of freedom for the pairing states are sublattice (orbital), represented by Pauli matrices $\tau_i$,  and spin, represented by Pauli matrices $\sigma_i$. 

The inversion operation $I$ flips momentum and interchanges the sublattices, $I=\tau_x  (\vec{k}\rightarrow -\vec{k})$. The mirror plane symmetry operators $M_x$ and $M_y$ are defined as usual, $M_j = \tau_0\otimes i\sigma_j (k_j\rightarrow -k_j)$. Since the sublattices in UTe$_2$ are aligned along the $z$ axis, $M_z$ is defined as $M_z = \tau_x\otimes i\sigma_z (k_z\rightarrow -k_z)$. The odd-parity basis functions belonging to each irreducible representation of this symmetry group are shown in Table \ref{tab:irrep_table}.  The basis functions are of the form $\tau_j\otimes (\vec{d}(k)\cdot \vec{\sigma})(i\sigma_y)$ for spin triplet states or $\tau_j\otimes(i\sigma_y)$ for spin singlet states ($j=0,x,y,z$). Generically, the gap function will be related to the basis functions listed here through a factor of the gap magnitude. 

\begin{table}[h]
    \centering
    \begin{tabular}{c|cccc|ccc}
    IR  & $I$ & $M_x$ & $M_y$ & $M_z$  & $\vec{d}(k)$ &$\tau$ \\
    \hline
     $A_u$    &-1 & -1 & -1 & -1 & $k_x\hat{x}$, $k_y\hat{y}$, $k_z\hat{z}$ & $\tau_x$, $\tau_0$ \\
     & & & & & $\hat{z}$ &  $\tau_y$\\
     $B_{1u}$  &-1  &1 &1 &-1  &  $k_y\hat{x}$, $k_x\hat{y}$ & $\tau_x$, $\tau_0$ \\
      & & & & & - &  $\tau_z$\\
     $B_{2u}$   &-1 &1 & -1 & 1  & $k_z \hat{x}$, $k_x\hat{z}$ & $\tau_x$, $\tau_0$ \\
      & & & & & $\hat{x}$ &  $\tau_y$\\
     $B_{3u}$  &-1 & -1 &1 &1   & $k_z\hat{y}$, $k_y\hat{z}$ & $\tau_x$,$\tau_0$ \\
      & & & & & $\hat{y}$ &  $\tau_y$\\
    \end{tabular}
    \caption{Classifications of odd parity states for $D_{2h}$ symmetry (orthorhombic crystal). The basis functions are of the form $\tau_j\otimes (\vec{d}(k) \cdot \vec{\sigma}) (i\sigma_y)$ with $j=0,x,y$; absence of a listed $\vec{d}(k)$ indicates a spin singlet state. Momentum labels ($k_i$) are symmetry labels and do not indicate the form of the functional dependence. }
    \label{tab:irrep_table}
\end{table}

With a finite magnetic field aligned along the crystalline $b$ axis, the orthorhombic symmetry is broken down to $C_{2h}^y$, as mirror symmetries along the $x$ and $z$ axes are destroyed. The irreducible representations in Table \ref{tab:irrep_table} are allowed to mix, distinguished now only by their behavior under $M_y$, as shown in Table \ref{tab:irrep_table_field}. 

\begin{table}[h]
    \centering
    \begin{tabular}{c|ccc}
    IR (In field)& IRs (Zero-field)  & $I$ & $M_y$  \\
    \hline
    $A_u$ & $A_u$, $B_{2u}$ & -1 & -1 \\
    $B_u$& $B_{1u}$, $ B_{3u}$ & -1 & 1
    \end{tabular}
    \caption{Classifications of odd parity states for $C_{2h}^y$ symmetry (orthorhombic crystal in finite $b$-axis magnetic field). States originally distinguished by their behaviors under $M_x$ and $M_z$ are now allowed in the same symmetry classification upon the breaking of $M_x$ and $M_z$ symmetries.}
    \label{tab:irrep_table_field}
\end{table}

\subsection{Tight binding model for UTe$_2$}

We now adopt a minimal two-band tight-binding model with one local orbital per uranium atom, which was previously established in Ref.~\onlinecite{Shishidou2021} and possesses the essential properties of sublattice structure and orthorhombic symmetry. This model captures all of the possible pairing symmetries: 
\begin{align}
h_t &= [\epsilon_0(k)-\mu] \tau_0\otimes \sigma_0  + f_{Ag}(k) \tau_x\otimes\sigma_0 \notag\\&+ f_{z}(k) \tau_y\otimes\sigma_0 + f_{y}(k) \tau_z\otimes\sigma_x\notag \\&+ f_{x}(k) \tau_z\otimes\sigma_y + f_{Au}(k) \tau_z\otimes\sigma_z.  
\label{eq:ht}
\end{align}
Here, $\tau_i$ are Pauli matrices on the orbitals (sublattices), and $\sigma_i$ are Pauli spin operators. The first three terms, with coefficients $\epsilon_0(k)$, $f_{Ag}(k)$, and $f_z(k)$ describe the kinetic energy of the itinerant electrons on the uranium atoms. They have the forms
\begin{align}
    \epsilon_0(k) &= t_1\cos k_x + t_2\cos k_y \notag\\
    f_{Ag}(k) &= m_0 + t_3 \cos(k_x/2)\cos(k_y/2)
\cos(k_z/2)\notag\\
    f_z(k) &= t_z \sin(k_z/2)\cos(k_x/2)\cos(k_y/2).
    \label{eq:kinetic}
\end{align}
The magnitude of each hopping integral ($t_1$, $t_2$, $m_0$, $t_3$, $t_z$) was found from DFT \cite{Shishidou2021}, and the precise values used in our work are listed in Sec. \ref{sec:tb_app}. Since the in-plane hopping in the $\hat{x}$ direction, $t_1$, is the largest kinetic energy scale, we set this as the unit of energy ($t_1 = 1$). 

The last three terms of Eq. \ref{eq:ht} are anisotropic spin-orbit couplings with momentum dependence given by
\begin{align}
     f_y(k)&= t_y\sin(k_y)\notag\\ 
     f_x(k)&= t_x\sin(k_x) \notag \\ 
     f_{Au}(k)&= t_u \sin(k_x/2)\sin(k_y/2)
\sin(k_z/2).
 \label{eq:soc}
\end{align}
 We anticipate that the scale of spin-orbit coupling is dictated by the geometry of the system as well. The inter-atomic distance between uranium atoms at different sites is smallest in the $a$ ($\hat{x}$) direction and largest along the diagonal connecting the body-centered site to the corners. Thus, throughout this work, we consider $t_x>t_y>t_u$. 


On top of the kinetic energy, we introduce a magnetic field $B$, coupled to  the spin via a Zeeman term,
\begin{equation}
h_B = -\tau_0\otimes (\vec{B} \cdot \vec{\sigma}) .
\end{equation}
We will mainly consider magnetic fields aligned along the crystalline $b$ axis, $\vec{B} \parallel \hat{y}$. The full Hamiltonian is then
 \begin{equation}
 h(k) = h_t - B\tau_0\otimes \sigma_y.  \label{eq:ham}
 \end{equation}

\section{Pairing Symmetries of Low- and High Field Phases \label{sec:pairing}}
\subsection{Superconducting susceptibility\label{sec:chi}}
Here, we evaluate the superconducting instabilities of the normal state via the pair field susceptibility $\chi$ in the absence of pairing interactions. Since experimental signatures of UTe$_2$ strongly suggest odd-parity superconductivity, we will consider only the instabilities towards inversion-odd pairing states (those listed in Table \ref{tab:irrep_table}). This approach reveals the odd-parity superconducting state favored by the band structure, as opposed to that favored by a specific interaction. 

Within mean field theory, the susceptibility of the normal state to a specific pairing channel $\Gamma$ is the linear response function of the normal state to the pairing ``field" with symmetry $\Gamma$. The susceptibility $\chi_\Gamma$ thus quantifies how easily the normal state forms pairs with symmetry $\Gamma$. Assuming that superconductivity arises from a weak-coupling instability in this system, the pairing channel for which $\chi_\Gamma$ is maximal determines the true superconducting order. 

To compare the susceptibilities to all pairing channels in UTe$_2$, we construct the superconducting susceptibility matrix $\chi$ with entries
\begin{align}
\chi_{ij} &= -\frac{1}{\beta} \sum_{\omega_n} \sum_{\vec{k}'}  Tr\bigg[D^\dagger_i(\vec{k}')  G(\vec{k}', i\omega_n) D_j(\vec{k}')\notag\\
&\quad \quad \quad \times G(-\vec{k}', -i\omega_n) \bigg],
\label{eq:susc}
\end{align} 
where $\omega_n$ are Matsubara frequencies,  $D_i(k)$ and $D_j(k)$ are basis functions of the orthorhombic symmetry (as listed in Table \ref{tab:irrep_table}), $G(k,i\omega)$ is the normal-state single-particle Green's function, and the sum over $k'$ is taken over the Fermi surface (as defined by Eq. \ref{eq:ham}). The diagonal entries $\chi_{ii}$ are the susceptibilities to forming a gap proportional to $D_i(k)$, in response to a pairing field with the same structure $D_i(k)$. Cross terms $\chi_{ij}$ (for $i\neq j$) are  the susceptibilities to forming a gap with structure $D_i(k)$, in response to a  pairing field of a \textit{a different} form $D_j(k)$. 
\begin{figure}
    \centering
    \includegraphics{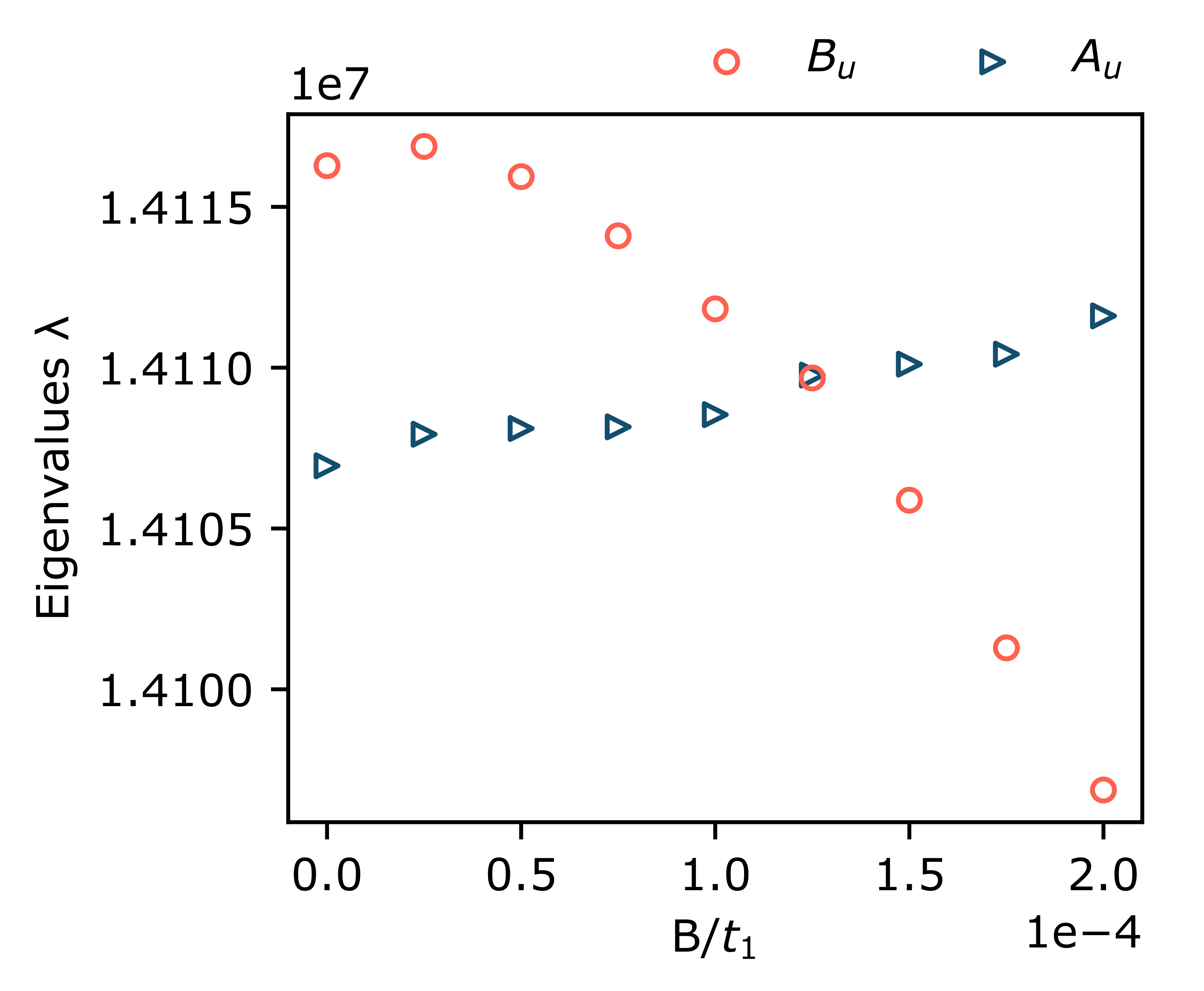}
    \caption{The two largest eigenvalues of the pair field susceptibility matrix $\chi$ as a function of applied magnetic field strength $B$, for a field aligned along the crystalline $b$-axis. This is found at a temperature $T=10^{-4}t_1$ and with approximately $6\times 10^4$ points on the Fermi surface. The dominant basis functions of type $B_u$ are $\tau_0\otimes( k_x \sigma_y) (i\sigma_y)$ and $\tau_x \otimes( k_x \sigma_y) (i\sigma_y)$, while the dominant basis functions of type $A_u$ are $\tau_0 \otimes( k_x \sigma_x) (i\sigma_y)$ and $\tau_x\otimes( k_x \sigma_x) (i\sigma_y)$. At around $B = 1.25\times10^{-4}t_1$, there is a crossing between the largest eigenvalues, indicating a transition between pairing states. }
    \label{fig:eigenvalues}
\end{figure}

Generically, if $D_i$ and $D_j$ belong to the same irreducible representation, $\chi_{ij}$ can be nonzero. Thus, the correct susceptibilities to compare are not those between the different basis functions but instead those between different eigenstates of $\chi$, which are mixtures of  basis functions in the same irreducible representation. The eigenvalues of $\chi$ are still a proxy for the logarithm of the superconducting transition temperatures $T_c$, and the true superconducting order has the form of the eigenvector corresponding to the largest eigenvalue.

Fig. \ref{fig:eigenvalues} shows the evolution of the largest eigenvalues of $\chi$ (Eq. \ref{eq:susc}) as a function of the applied magnetic field strength $B$. The eigenvalues are labeled by the symmetry classifications that their corresponding eigenvectors belong to. At low fields, states in the $B_u$ classification are favored, and spin-triplet states with  $\vec{d}\parallel \hat{y}$ ($\tau_x\otimes k_x \sigma_y (i\sigma_y)$ and $\tau_0\otimes k_x \sigma_y (i\sigma_y)$) dominate. At a critical field strength of $B_c \sim 10^{-4} t_1$, there is a crossing of the largest eigenvalues, signalling a transition from $B_u$ to $A_u$. The dominant basis functions at high field are spin-triplet with  $\vec{d}\parallel \hat{x}$ ($\tau_x\otimes k_x \sigma_x (i\sigma_y)$ and $\tau_0\otimes k_x \sigma_x (i\sigma_y)$).  

This level crossing may be understood as a result of the competition between the magnetic field and spin-orbit coupling. At $B=0$, the spin-orbit coupling largely determines the pairing state to be in $B_u$, with a primary spin component in the $ac$ plane. Since this is energetically unfavorable in the presence of a magnetic field along the crystalline $b$ axis, increasing the magnetic field strength ultimately overwhelms the spin-orbit coupling and drives a transition to a $A_u$.  

While the qualitative picture offered here is consistent with the experimental observations, we find that the relative splitting between eigenvalues (effective differences in $T_c$) between the two phases is very small. This may be an indication that the competition between spin-orbit coupling and the magnetic field is insufficient to fully explain the transition.
In the next section, we consider the effects of ferromagnetic pairing interactions and assess the robustness of the results described above. 


\subsection{Self-consistent mean-field approach\label{sec:Delta}}
Our analysis of the normal state instabilities suggests that, even without considering any  specific pairing interactions, there is a tendency towards a transition between superconducting states with distinct pairing symmetries due to competition between spin-orbit coupling and applied magnetic field. We now account for interactions and determine the pairing states favored by a potential, rather than the kinetic, energy; we identify the nature (first- or second-order) of the transition between pairing symmetries and assess the effect of interactions on the value of the critical field $B_c$. 

We consider an on-site, opposite-sublattice ferromagnetic interaction 
\begin{equation}
H_I = -\left(\sum_{i} J_x S_{i,1}^x S_{i,2}^x +  J_y S_{i,1}^y S_{i,2}^y + J_z S_{i,1}^z S_{i,2}^z\right),\label{eq:ferroint}
\end{equation}
where $i$ is a site index and $1,2$ are sublattice indices. 
While there are a plethora of other conceivable local interactions, we choose interactions of this particular form, as suggested by DFT calculations\cite{Xu2019} and supported by neutron scattering experiments \cite{Knafo2021}. Note that the results of Ref. \onlinecite{Knafo2021} suggest nearest neighbor antiferromagnetic interactions along the $y$ axis (favoring singlet pairing) and nearest-neighbor ferromagnetic interactions along the $x$ axis (favoring the states identified in Sec. \ref{sec:chi}) \textit{in addition }to the on-site ferromagnetic interactions we have chosen here. However, recent work incorporating this more general form of the interaction and a different normal-state Hamiltonian\cite{Chen2021} identifies the same zero-field state as we do, indicating that the results reported here are likely robust to these additional interactions.




To find the gap function $\Delta(k)$ in the presence of these interactions, we take a standard mean-field approach, decoupling the four-fermion interaction and defining the gap function in terms of the interaction. In the spin-orbit Nambu basis, the Bogoliubov deGennes (BdG) Hamiltonian takes the form
\begin{equation}
    H_{BdG} = \begin{pmatrix}
    h(k) & \Delta(k) \\ 
    \Delta^\dagger(k) & -h(-k)^T
    \end{pmatrix},
    \label{eq:hbdg}
\end{equation}
where $h(k)$ and $\Delta(k)$ are $4\times 4$ matrices,
and has eigenvectors and eigenvalues satisfying
\begin{equation}
    H_{BdG} \begin{pmatrix}
    \{u^n_{k\mu}\}\\
    \{v^n_{-k\mu}\} 
    \end{pmatrix} = E_{kn} \begin{pmatrix}
    \{u^n_{k\mu}\}\\
    \{v^n_{-k\mu}\}
    \end{pmatrix}. 
    \label{eq:bdg_eig}
\end{equation}

The mean-field self-consistency condition for $\Delta(k)$ is then
\begin{align}
\Delta_{\mu\nu}(\vec{k}) &= \sum_{\vec{k}'} \sum_{\mu' \nu'} V_{\mu\nu\mu'\nu'} (\vec{k},\vec{k}') \notag\\
&\times\sum_n \left[(u_{\vec{k'\mu_3}})^{n} (v^*_{-\vec{k'\mu_4}})^{n} \tanh\left(\frac{E_{\vec{k}' n}}{2k_B T} \right)\right],
\label{eq:gapeq}
\end{align}
where $V_{\mu\nu\mu'\nu'}$ is the pairing interaction generated from Eq. \ref{eq:ferroint} (see Sec. \ref{sec:bcs_app}), the sum over $\vec{k}'$ is over the Fermi surface, $\mu'$ and $\nu'$ are generalized spin-orbit indices, and $E_{kn}$, $u_{k\mu}^n$, and $v_{-k\mu}^n$ are defined by Eq. \ref{eq:bdg_eig}. In contrast to our approach in Sec. \ref{sec:chi}, we do not assume a particular parity of the gap function. Instead, the solutions $\Delta(k)$ to Eq. \ref{eq:gapeq} are generically admixtures of the basis functions in the orthorhombic symmetry group which are allowed by symmetry ($I$, $M_i$) to mix. Thus, the following results reveal which states are favored by the pairing interactions of Eq. \ref{eq:ferroint}, under the symmetry constraints provided by the normal-state Hamiltonian (Eq. \ref{eq:ham}). 
\begin{figure}[h]
    \centering
    \includegraphics{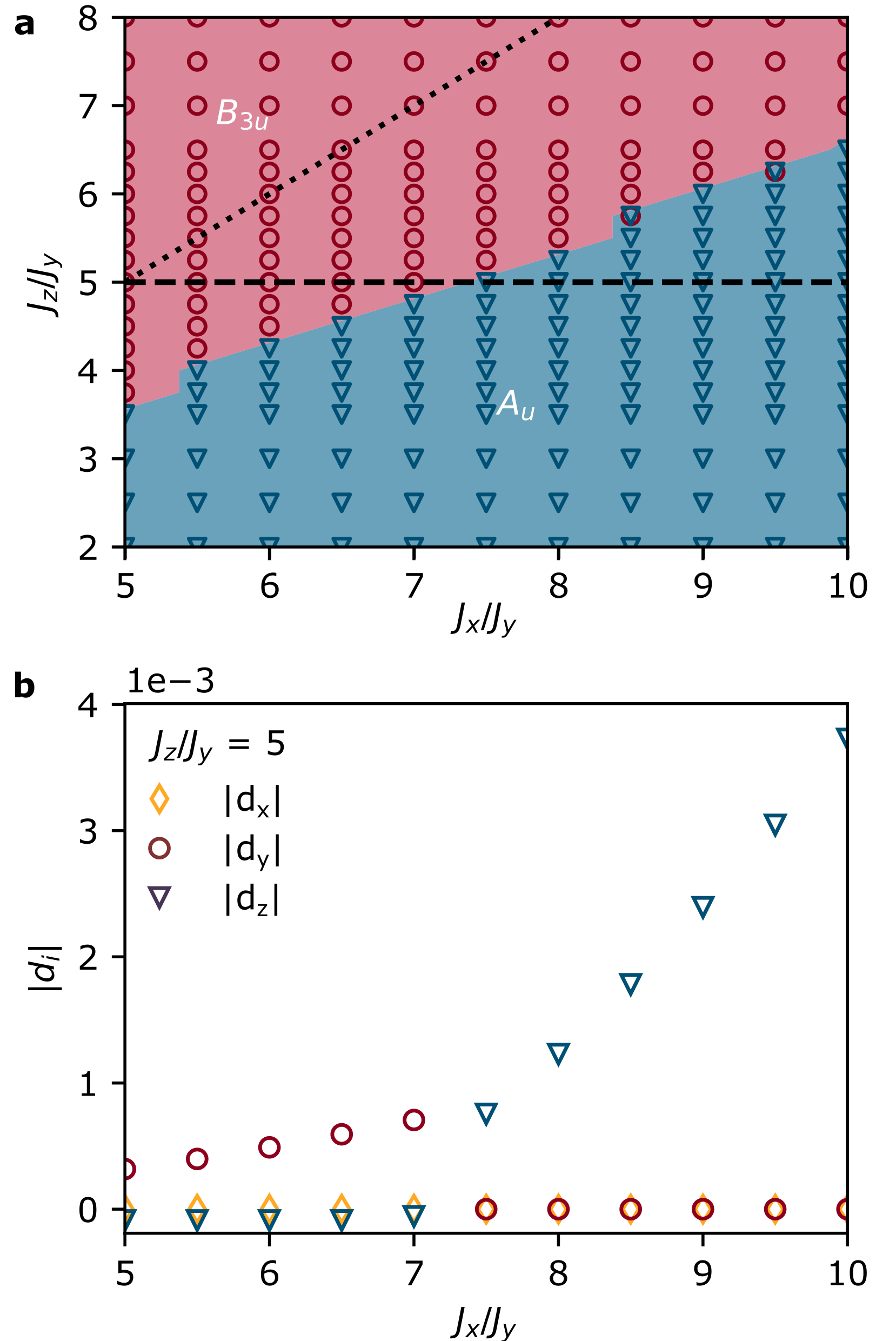}
    \caption{ (a) The zero-field, zero-temperature ($B=0$, $T=0$) phase diagram in interaction parameter space obtained from solving the self-consistent gap equation numerically for $J_y = 10^{-3}t_1$. Solutions are of the form $\Delta = \tau_y\otimes (\vec{d}\cdot \vec{\sigma})(i\sigma_y)$ in the spin-orbit basis, with $\vec{d} = \vec{d}(J_x, J_y, J_z)$. The dotted line denotes the separation between realistic ($J_x>J_z>J_y$) and unrealistic parameters, while the dashed line indicates the value of $J_z$ for the line cut plotted in Panel (b). (b) A representative line-cut in the space of parameters, showing a first-order transition between $B_{3u}$ and $A_u$ as a function of $J_x/J_y$. Shown are the magnitudes of $d_i$ ($i=x,y,z$) at $J_z/J_y = 5$, which exhibit discontinuous jumps indicating a first-order transition between pairing states.}
    \label{fig:pd}
\end{figure}

\begin{figure}
    \centering
\includegraphics{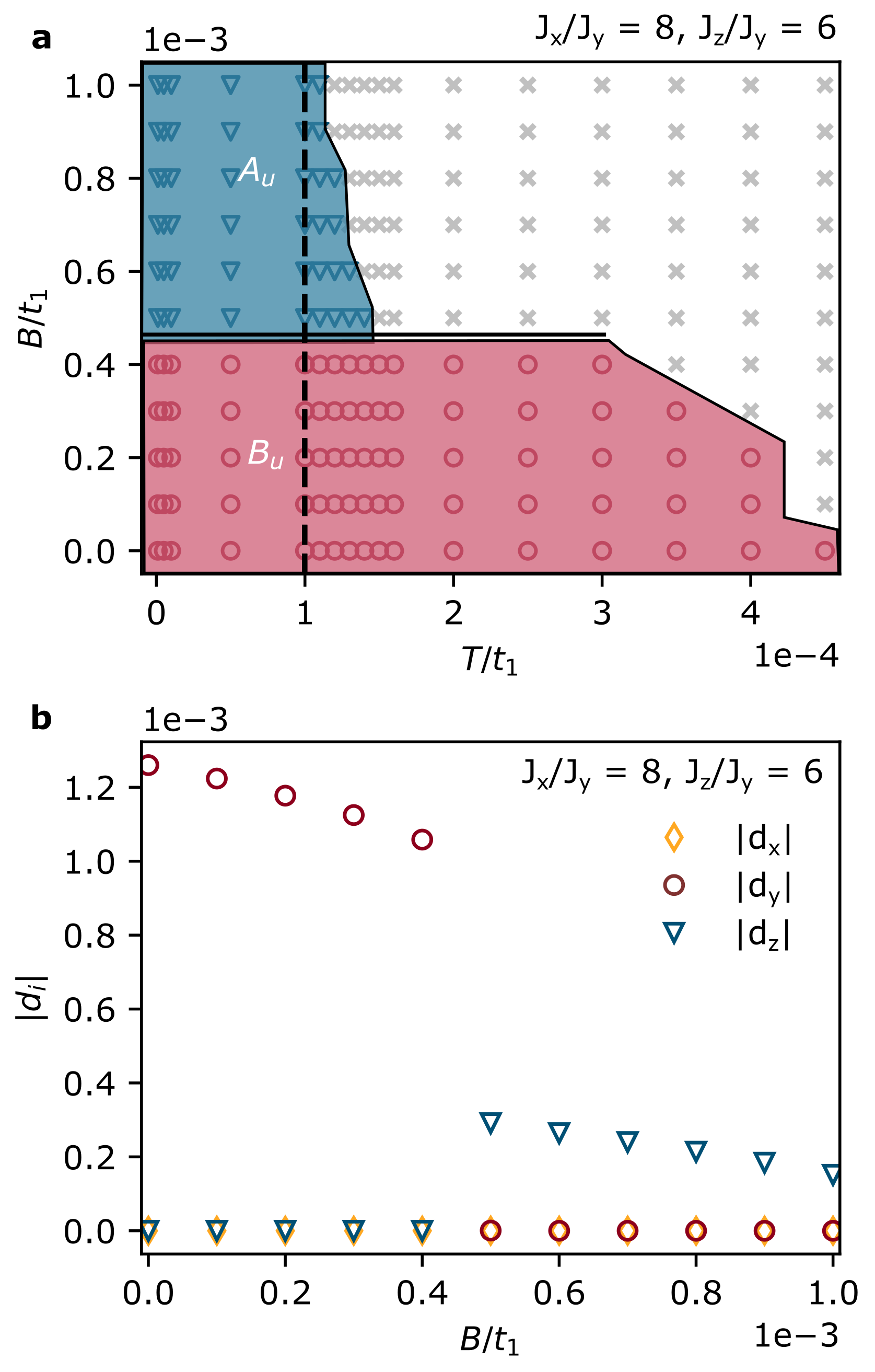}
    \caption{Solutions to the gap equation with interaction parameters $J_x/J_y = 8$, $J_y=10^{-3}t_1$, $J_z/J_y = 6$.  (a) The phase diagram in temperature $T$ and  magnetic field strength $B$ for a $\vec{B} = B\hat{y}$, as found by solving the gap equation (Eq. \ref{eq:gapeq}). In the red region, the superconducting gap function has the form $\tau_y\otimes d_y\sigma_y (i\sigma_y)$, while in the blue region, the gap function has the form $\tau_y\otimes d_z\sigma_z (i\sigma_y)$. The unshaded region indicates the region in which the gap function vanishes ($d_x,d_y,d_z<\epsilon$, for $\epsilon$ the error tolerance $\epsilon=10^{-5} t_1$), such that the system is in the normal state. The temperature-induced transitions are second-order (single line) while the field-induced transitions  are first-order (double line). A dashed line at $T=10^{-4}t_1$ indicates the line cut examined in Panel (b).
    (b) Gap magnitudes $|d_i|$ ($i=x,y,z$) as a function of applied magnetic field strength $B$, for a field aligned along the crystalline $b$-axis at temperature $T=10^{-4}t_1$. At a field strength of approximately $B/t_1\sim 5\times 10^{-4}$, there is a first-order transition from the $B_{u}$ pairing state at low fields to the $A_u$ pairing state at high fields. }
    \label{fig:TvB}
\end{figure}

We solve Eq. \ref{eq:gapeq} by iteration, starting with a random initial matrix $\Delta_0(\vec{k})$. We consider the solution to be converged after $n$ iterations when $\Delta_n$ and $\Delta_{n-1}$ satisfy the convergence condition $|\Delta_n(\vec{k})-\Delta_{n-1}(\vec{k})| < 10^{-8}t_1 \approx 1\text{ neV}$. Details of the this procedure are in Sec. \ref{sec:bcs_app}. 

While we allow for both even and odd parity solutions of Eq. \ref{eq:gapeq}, we have found that all non-trivial solutions have odd parity and take the form $\Delta = \tau_y\otimes (\vec{d}\cdot \vec{\sigma})(i\sigma_y)$, where  $\vec{d}$ is momentum-independent. For the remainder of this section, we will refer to the gap function by the orientation of $\Vec{d}$, always implicitly assuming the form $\Delta = \tau_y\otimes (\vec{d}\cdot \vec{\sigma})(i\sigma_y)$. We also absorb the gap magnitude into $\vec{d}$, such that $|\Delta|^2 = |\vec{d}|^2$.   
In the same spirit as Ref. \onlinecite{Chen2021}, we will allow for anisotropy in the interactions. We first consider how $J_x$, $J_y$, and $J_z$ determine the pairing symmetries at zero-field ($B=0$) and identify realistic values for the exchange energies. 

Fig. \ref{fig:pd}a is the phase diagram in the space of interaction parameters $J_x/J_y$ and $J_z/J_y$, for $B=0$, $T=0$, and $J_y = 10^{-3} t_1$. At $B=0$,  $\vec{d}\parallel \hat{x}$, $\vec{d}\parallel \hat{y}$, and $\vec{d}\parallel \hat{z}$ belong to distinct symmetry classifications (Table \ref{tab:irrep_table}). The nature of the transitions between the pairing states as a function of the interactions $J_x$ and $J_z$ can be determined by analyzing the gap magnitudes. Fig. \ref{fig:pd}b shows the magnitudes of $d_i$ as a function of $J_x/J_y$ for a fixed $J_z/J_y = 5$. This reveals a first-order transition between $B_{3u}$ and $A_u$ in the interaction parameter space at $B=0$.

The actual interactions present in UTe$_2$ are modeled well only by a region of the interaction space shown in Fig. \ref{fig:pd}a. Since the zero-field state is experimentally known to be suppressed by a $b$-axis magnetic field, we identify the $B_{3u}$ state ($\vec{d}\parallel \hat{y}$) as a good candidate for the low-field state, as it is a triplet state with spin in the $ac$ plane. As shown in Fig. \ref{fig:pd}a, the $B_{3u}$ state is favored for interactions $J_x,J_z> J_y$. The parameter range in which we find a $B_{3u}$ state is consistent with expectations from the magnetic properties of the normal state.  Since the $a$ axis is the easy magnetic axis, and the $b$ axis is the hard magnetic axis \cite{Ran2019}, the physical interaction parameters are likely $J_x > J_z > J_y$. The separation between the realistic and unrealistic interaction parameter regimes is shown in Fig. \ref{fig:pd}a as a dotted line.

So far, we have solved the self-consistency equation (Eq. \ref{eq:gapeq}) for the gap function at $B=0$ and $T=0$, for a variety of interactions. We now consider the behavior of the gap at finite temperature and magnetic field and construct a phase diagram in the space of temperature $T$ and applied field $B$ (along the $\hat{y}$ axis), shown in Fig. \ref{fig:TvB}a. Specifically, we investigate the nature of the transition between the $B_u$ low-field state and the $A_u$ high-field state as a function of applied field strength, as shown in Fig. \ref{fig:TvB}b.

 We choose a representative set of parameters corresponding to a gap in the phase $B_{3u}$ ($\Vec{d}\parallel \hat{y}$) at $B=0$ and $T=0$:  $J_x/J_y = 8$, $J_z/J_y=6$, and $J_y = 10^{-3}t_1$. We then solve the self-consistent gap equation (Eq. \ref{eq:gapeq}) with $B> 0$ and $T>0$. As shown in Fig. \ref{fig:TvB}a, at a finite field $B>0$, there is a first-order transition between the pairing states with $\Vec{d}\parallel \hat{y}$ and $\Vec{d}\parallel\hat{z}$, while finite temperature  transitions between the superconducting states and the normal state remain second-order. In principle, the high-field $A_u$ state could be an mixture of the $\vec{d}\parallel \hat{x}$ and $\vec{d}\parallel{z}$ states (see Table \ref{tab:irrep_table_field}), but we find that any $\vec{d}\parallel \hat{x}$ is suppressed by the large value of $J_x$.
 
 Fig. \ref{fig:TvB}b shows the evolution of $\vec{d}$ as a function of the applied magnetic field strength $B/t_1$. Upon increasing $B$, the $B_{u}$ state is  suppressed, and we observe a first-order transition to $A_u$ pairing symmetry at around $B/t_1=5\times 10^{-4}$. The transition is between the same symmetry classifications as those identified in Sec. \ref{sec:chi}, but the transition occurs at an enhanced critical field $B_c$, which may be attributed to the cooperation between spin-orbit coupling and interactions to  stabilize of the low-field state. 

\section{Properties of the Low- and High- Field Phases \label{sec:properties}}

\subsection{Sensitivity to angle \label{sec:angle}}
The low- and high-field superconducting phases in UTe$_2$ are distinguished by their sensitivity (or lack thereof) to the angle of the applied magnetic field with respect to the crystalline $b$ axis; the high-field phase is sensitive to the angle of the field, whereas the low-field phase is not. We now show that our results are consistent with this observation through a qualitative argument and by providing numerical evidence in support of this claim. 

From both the analysis of the superconducting susceptibility and of the mean-field solution in the presence of ferromagnetic interactions, we find that the low-field phase is a triplet state with primarily $\vec{d}\parallel \hat{y}$, whereas the high-field phase has primarily $\Vec{d}\parallel \hat{x}$ or $\Vec{d}\parallel \hat{z}$. 
\begin{figure}[h]
    \centering
\includegraphics[scale=0.25]{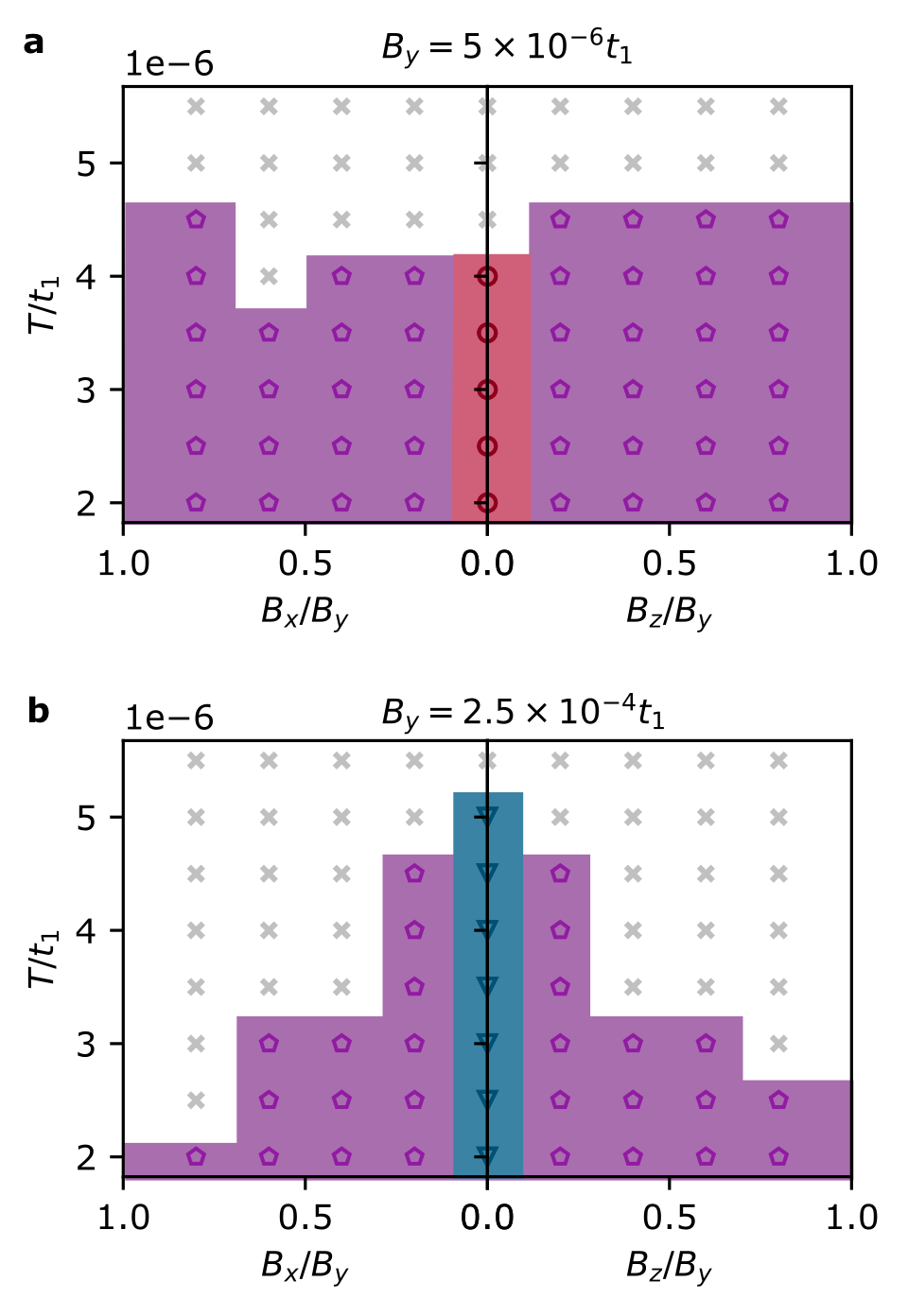}
    \caption{Phase diagrams in temperature $T$ and magnetic fields $B_x$ or $B_z$, for solutions of the gap equation with interaction parameters $(J_x, J_y, J_z)=(2.8\times10^{-3}t_1, 1.5\times 10^{-3}t_1,2.6\times10^{-3}t_1)$ for (a) $B_y= 5\times 10^{-6}t_1$ and (b) $B_y= 2.5\times 10^{-4}t_1$. At $B_x=B_z=0$, the solution at (a) $B_y= 5\times 10^{-6}t_1$ is $\tau_y\otimes (d_z\sigma_y)(i\sigma_y)$ (red shaded region, circles), and the solution at (b) $B_y= 2.5\times 10^{-4}t_1$ is $\tau_y\otimes (d_z\sigma_z)(i\sigma_y)$  (blue shaded region, triangles). In the presence of a field misaligned with the crystal axes, all mirror symmetries are broken, and there is only one irreducible representation (purple shaded region, pentagons). The unshaded region indicates the region in which the gap function vanishes ($d_x,d_y,d_z<\epsilon$, for $\epsilon$ the error tolerance $\epsilon= 1\times 10^{-6} t_1$), such that the system is no longer superconducting. }
    \label{fig:fieldtilt}
\end{figure}
Since the spin in a triplet state is proportional to $\Vec{d}\times \Vec{d}^*$, a  state with $\vec{d}\parallel \hat{y}$ (spin in the $xz$ plane) will be suppressed in the presence of a large magnetic field along the $\hat{y}$ direction. However, such a state should be relatively insensitive to a field in the $\hat{x}$ or $\hat{z}$ directions. This is consistent with the experimental results in the low-field phase. In contrast, the high-field $A_u$ phase with $\Vec{d} =  d_x\hat{x}+d_z\hat{z}$ is stable to large fields along $\hat{y}$ but is suppressed by fields along $\hat{x}$ or $\hat{z}$. More concretely, the suppression of a given pairing state by a time-reversal symmetry breaking perturbation may be quantified by the field fitness function \cite{Ramires2016, Ramires2018, Cavanagh2022}. For example, spin-orbit coupling determines how severely the specific high-field solution $\tau_y\otimes (d_z\sigma_z)(i\sigma_y)$ found in Sec. \ref{sec:Delta} is suppressed by tilting of the field in the $a$ and $c$ directions (see Sec. \ref{app:field_fitness}). 

We also explicitly demonstrate the responses of states identified in Sec. \ref{sec:Delta} to magnetic fields off of the $b$ axis. Specifically, we again solve the gap equation (Eq. \ref{eq:gapeq}) at various temperatures $T$ and with interactions $(J_x, J_y, J_z)=(2.8\times10^{-3}t_1, 1.5\times 10^{-3}t_1,2.6\times10^{-3}t_1)$. We consider two fixed values of $B_y$, corresponding to the low-field state at $\vec{B}=(0,5\times 10^{-6}t_1, 0)$ and the high-field state at $\vec{B}=(0,2.5\times 10^{-4}t_1, 0)$ and introduce finite $B_x$ and $B_z$. As shown in Fig. \ref{fig:fieldtilt}, the critical temperature of the low-field state does not change significantly upon the introduction of $B_x$ or $B_z$. In contrast, the critical temperature of the high-field state decreases with increasing $B_x$ or $B_z$. Qualitatively, this matches the experimental observations\cite{Rosuel2022, Kinjo2022}.

\subsection{Nodal structure}
We now consider the nodal structure of the gap functions found in the self-consistent calculation. Calorimetric measurements\cite{Kittaka2020, Metz2019, Ran2019, Bae2021}, magnetic penetration depth measurements \cite{Ishihara2022},  and NMR $1/T_1$ relaxation rate measurements \cite{Nakamine2019} in UTe$_2$ at zero-field show evidence of point nodes in the superconducting gap function, but there is no global consensus on the locations of these nodes. From the self-consistent solution to Eq. \ref{eq:gapeq} in zero field ($B=0$), we find that the candidate zero-field ($B_{3u}$) phase has point nodes along the $a$ axis. 

In a one-band system, the nodal structure of a gap can be deduced from its symmetry classification. This is because the momentum dependence of the basis functions dictates the nature of the nodes (none, point, or line). However, in a multiband system, the nodal structure cannot be straightforwardly related to the symmetry classification of the gap, as the basis functions gain nontrivial structure in the band basis \cite{Agterberg2017,Agterberg2017a,Zhou2008}.

We thus identify the nodal structure of the solution to Eq. \ref{eq:gapeq} by projecting $\Delta$ into the band basis. To simplify our calculation, we consider the projection only at $k_z=0$ ($f_z = f_u = 0$) (see Sec. \ref{sec:nodes_app}). For a given pairing state, the stable nodes are those which are not suppressed by adding an arbitrary mixture of other pairing states within the same symmetry classification. The projection at $k_z=0$ yields that the candidate zero-field ($B_{3u}$) phase has stable point nodes along the $a$ axis. These results are consistent with transport measurements \cite{Metz2019} which identify point nodes in the $ab$ plane and field-angle-resolved measurements of the specific heat \cite{Kittaka2020} which identify point nodes along the $a$ axis for the zero-field superconducting state. 

\section{Discussion}
 In this work, we have determined the pairing symmetries of the low- and high-field phases of UTe$_2$ within mean field theory for a minimal Hamiltonian. We find that the field-induced transition between pairing symmetries in UTe$_2$ is a transition from states of the $B_u$ classification at low fields to states of the $A_u$ classification at high fields. These pairing states are consistent with the experimental signatures within each phase, namely the suppression of the high-field phase upon tilting the magnetic field away from the $b$ axis and the lack thereof in the low-field phase. Furthermore, our predictions of the nodal structure of the gap at zero-field are consistent with the results of thermal transport measurements. 

 However, the $T-B$ phase diagram found in our calculations (Fig. \ref{fig:TvB}a) does not reflect the phenomenon of reentrant superconductivity. This suggests that fluctuations, which are neglected in our mean-field approach, may be responsible for an increase of $T_c$ with increasing $B$ in the high-field phase. The details of reentrant superconductivity and other the phenomena in UTe$_2$ are also undoubtedly influenced by factors such as disorder\cite{Rosa2022}, vortex formation\cite{Iguchi2022}, and orbital-field coupling, which we have also neglected. Even so, the minimal model used here, which includes only the most essential structural elements of UTe$_2$, captures the field-driven transition between different pairing symmetries and qualitative signatures of these high- and low-field phases at low temperatures. This suggests that the orthorhombic crystal symmetry and sublattice structure of UTe$_2$ play the largest roles in determining its superconducting states. 

While our approach using itinerant electrons successfully determines the nature of superconductivity in UTe$_2$, the underlying \textit{mechanisms} behind superconductivity remain unexplained.  Specifically, we assume here a particular form for the local ferromagnetic interaction (Eq. \ref{eq:ferroint}) and a specific hierarchy of the anisotropic spin-orbit couplings. A derivation of the interaction and SOC are beyond the scope of this work, but we acknowledge that symmetry principles can justify the general form of the model and the anisotropic nature of the interactions and SOC but cannot fully explain the origins of these terms. Below, we conjecture how a complementary perspective may supply a satisfactory conceptual understanding.

UTe$_2$ is found to have signatures of a strong-coupling superconductor, and agreement between DFT and experiments depends sensitively on the Hubbard interaction parameter\cite{Aoki2022a}, suggesting that correlation effects in UTe$_2$ are essential. Thus, the more fundamental questions about the origins of interactions and mechanisms responsible for superconductivity may be better answered from a perspective of local, microscopic physics complementary to the one presented here. More generally, a complete description of the phenomena in UTe$_2$ and related heavy-fermion superconductors likely requires a combination of both the local and itinerant  perspectives.

\section{Acknowledgements}
 We thank S. Brown, T. Hazra for helpful discussions. JJY was supported by the National Science Foundation Graduate Research Fellowship under Grant No. DGE-1656518. SR was supported in part by the US Department of
Energy, Office of Basic Energy Sciences, Division of Materials Sciences and Engineering, under contract number
DE-AC02-76SF00515. DFA and YY were supported by by the US Department of Energy, Office
of Basic Energy Sciences, Division of Materials Sciences
and Engineering under Award DE-SC0021971.

\bibliography{bibliography}
\newpage
\begin{widetext}
\section{Appendix \label{sec:App}}
\subsection{Parameters of the tight-binding model\label{sec:tb_app}}
The constants in Eq. \ref{eq:kinetic} and Eq. \ref{eq:soc} used in this work are listed here. For the calculation of the pair-field susceptibility and its eigenvalues and eigenstates (Sec. \ref{sec:chi}), we use $(t_1,  \mu, t_2, m_0, t_3, t_z, t_x, t_y, t_u) = (-1, 1.446, 0.76, -0.695, 0.83, -0.83, 0.785, 0.224, 0.112)$. For the solution of the self-consistent gap equation (Sec. \ref{sec:Delta}), we use $(t_1,  \mu, t_2, m_0, t_3, t_z, t_x, t_y, t_u) = (-1, 1.446, 0.76, -0.695, 0.83, -0.83, 0.448, 0.224, 0.112)$. In Sec. \ref{sec:angle}, we use $(t_1,  \mu, t_2, m_0, t_3, t_z, t_x, t_y, t_u) = (-1, 1.446, 0.76, -0.695, 0.83, -0.83, 0.15, 0.12, 0.11)$. 

\subsection{Solving the self-consistent gap equation with  ferromagnetic interactions\label{sec:bcs_app}}
Generically, a two-body interaction in real space takes the form $H_I = \sum_{i,j} \sum_{\mu,\nu,\mu',\nu'} U_{i}^{\mu \nu \mu' \nu'} c^\dagger_{i,\mu} c_{i,\mu'}c^\dagger_{j,\nu} c_{j,\nu'} $. Here, $i$ and $j$ are site labels, while $\mu('), \nu(')$ are generalized spin-orbit indices. Via a Fourier transform, one can always express this interaction in BCS form as
\begin{equation}
    H_I =\sum_{k,k'} V_{\mu_1,\mu_2,\mu_3,\mu_4}(\vec{k},\vec{k}') c_{k,\mu_1}^\dagger c_{-k,\mu_2}^\dagger c_{-k', \mu_3} c_{k', \mu_4},
\end{equation}
where $V$ is the effective BCS pairing interaction entering Eq. \ref{eq:gapeq} and repeated indices are summed over. 

For the local ferromagnetic interaction described in Eq. \ref{eq:ferroint}, 
\begin{align}
    V_{\mu_1,\mu_2,\mu_3,\mu_4}(\vec{k},\vec{k}') c_{k,\mu_1}^\dagger c_{-k,\mu_2}^\dagger c_{-k', \mu_3} c_{k', \mu_4} &= \sum_{a=x,y,z} J_a \sigma_a^{s's} \sigma_a^{p'p} c_{k,\tau=1, s'}^\dagger c_{-k,\tau=2, p'}^\dagger c_{-k', \tau=2, s} c_{k', \tau=1, p}.
\end{align}
for $\sigma_a$ Pauli matrices on the spins and $\tau$ indexing the sublattices. 

Then, the matrix $V$ is written: 
\begin{equation}   V_{\mu_1,\mu_2,\mu_3,\mu_4}(\vec{k},\vec{k}') = \epsilon^{\mu_1\mu_2} \epsilon^{\mu_3\mu_4}\sum_{a=x,y,z} \frac{J_a }{2}\left[\left(P^\tau_1 \sigma_a \right)^{\mu_1\mu_4}\left( P^\tau_2 \sigma_a \right)^{\mu_2\mu_3} + \left(P^\tau_2 \sigma_a \right)^{\mu_1\mu_4}\left( P^\tau_1 \sigma_a \right)^{\mu_2\mu_3}\right]
\label{eq:V_bcs}
\end{equation}

We use iteration to solve the self-consistent gap equation 
 Eq. \ref{eq:gapeq} with the interactions described by Eq. \ref{eq:V_bcs}. 

\subsection{Field-fitness functions for the solutions of the gap equation \label{app:field_fitness}}
While the numerical solutions to the gap equation show how the gap magnitudes of different pairing states evolve under a magnetic field, they do not offer insight as to what controls the suppression of a given pairing state by a given field. We quantify the pair-breaking effects of a magnetic field on the solutions of the self-consistent gap equation ($\tau_y\otimes \sigma_y(i\sigma_y)$ at low fields and $\tau_y\otimes \sigma_z(i\sigma_y)$ at high fields) by the field-fitness function as defined by Cavanagh \textit{et. al.} \cite{Cavanagh2022}. If the field-fitness function for a given pairing state and perturbation vanishes ($F_k=0$), then the perturbation does not have any depairing effects; on the other hand, $F_k=1$ indicates maximal pair-breaking. 

To understand the suppression of the high-field phase, we find the field-fitness functions for $\tau_y\otimes \sigma_z (i\sigma_y)$ in a magnetic field in the $\hat{x}$ or $\hat{z}$ directions. The result is:

\begin{align}
    F_k(B\parallel \hat{x})^{d\parallel z} &\propto \frac{f_{x}^2 f_{Ag}^2 }{(f_x^2+f_y^2+f_{z}^2)(f_z^2+f_{Ag}^2+f_{y}^2)} \\ 
    F_k(B\parallel \hat{z})^{d\parallel z} &\propto \frac{f_z^2(f_x^2+f_y^2+f_z^2+f_{Au}^2+f_{Ag}^2)}{(f_x^2+f_y^2+f_z^2)(f_z^2+f_{Au}^2+f_{Ag}^2)}
\end{align}

Generically, these will be nonzero over the Fermi surface, thus resulting in suppression of the high-field phase.

We compare these to the field-fitness functions of the low-field phase:  
\begin{align}
    F_k(B\parallel \hat{x})^{d\parallel y} &\propto \frac{f_{Au}^2 f_{Ag}^2 }{(f_z^2+f_y^2+f_{Au}^2)(f_z^2+f_{Ag}^2+f_{y}^2)} \\ 
    F_k(B\parallel \hat{z})^{d\parallel y} &\propto \frac{f_{y}^2 f_{Ag}^2 }{(f_z^2+f_y^2+f_{Au}^2)(f_z^2+f_{Ag}^2+f_{Au}^2)}
\end{align}

The kinetic energy scale is taken to be larger than the spin-orbit coupling energy scale: $f_z^2, f_{Ag}^2 > f_y^2, f_x^2, f_{Au}^2$.

Since the x-direction is the shortest bond, we expect that $f_x> f_{Au}$. This leads to $F_k(B\parallel x)^{d\parallel z} > F_k(B\parallel x)^{d\parallel y}$.

Additionally, $F_k(B\parallel z)^{d\parallel z} > F_k(B\parallel z)^{d\parallel y}$ for a similar reason. In summary, we argue here that the low-field $B_u$ phase is \textit{less} suppressed by fields in the $\hat{x}$ and $\hat{z}$ directions than the high-field $A_u$ phase; this agrees with experimental results and the claims in Sec. \ref{sec:properties}, and it provides some intuition for the terms responsible for pairing suppression. 


\subsection{Gap functions in the band basis \label{sec:nodes_app}}

In the band basis, the basis functions may be projected onto a single band. The momentum dependence of the gap in the band basis determines the nodal structure. Since basis functions of a given symmetry are able to mix, we identify nodes of a given symmetry as those which survive under arbitrary mixing of the basis functions. 

Tables \ref{tab:nodes_Bu} and \ref{tab:nodes_Au} list the momentum dependence of each basis function in Table \ref{tab:irrep_table}  found using simplified Hamiltonian at $k_z=0$. From the momentum dependence of the basis functions in each irreducible representation, we find that arbitrary mixtures of functions in the $B_{1u}$ classification have point nodes where  $k_x=0,\pm\pi$ and $k_y=0,\pm\pi$; mixtures of $B_{2u}$ have point nodes where $k_x=0,\pm \pi$ and $k_z=0,\pm\pi$; and mixtures of $B_{3u}$ have point nodes where $k_y=0,\pm\pi$ and $k_z=0,\pm\pi$.  Mixtures of the $A_u$ basis functions have no nodes generically.

\begin{table}
    \centering
    \begin{tabular}{c|cc}
 $\Delta$ (SO basis) & $D_{2h}$ Classification & Momentum dependence \\
    \hline
   $ \tau_0 k_y \sigma_x (i\sigma_y)$ & $B_{1u}$ & $\sin k_y\left( \frac{f_y}{\sqrt{f_x^2+f_y^2}}\hat{z} + \frac{f_{Ag}^2 f_x}{\sqrt{f_x^2+f_y^2}(f_{Ag}^2+f_x^2+f_y^2+\sqrt{(f_x^2+f_y^2)(f_{Ag}^2+f_x^2+f_y^2)}}\hat{y}\right)$\\
   $ \tau_0 k_x \sigma_y (i\sigma_y)$ & $B_{1u}$ & $\sin k_x\left(\frac{f_x}{\sqrt{f_x^2+f_y^2}}\hat{z} + \frac{f_{Ag}^2 f_y}{\sqrt{f_x^2+f_y^2}(f_{Ag}^2+f_x^2+f_y^2+\sqrt{(f_x^2+f_y^2)(f_{Ag}^2+f_x^2+f_y^2)})} \hat{x}\right)$ \\
   $ \tau_x k_y \sigma_x (i\sigma_y)$ & $B_{1u}$ & $\sin k_y\left(\frac{f_x}{\sqrt{f_x^2+f_y^2}}\hat{y} + \frac{f_{Ag}f_y}{\sqrt{f_x^2+f_y^2}(\sqrt{f_{Ag}^2 + f_x^2+f_y^2}+\sqrt{f_x^2+f_y^2})} \hat{z}\right) + \sin k_y \frac{f_{Ag} f_y}{f_{Ag}^2+f_x^2+f_y^2+\sqrt{(f_x^2+f_y^2)(f_{Ag}^2+f_x^2+f_y^2)}}$\\
   $ \tau_x k_x \sigma_y (i\sigma_y)$ & $B_{1u}$ & $\sin k_x\left(-\frac{f_y}{\sqrt{f_x^2+f_y^2}}\hat{y} + \frac{f_{Ag}f_x}{\sqrt{f_x^2+f_y^2}(\sqrt{f_{Ag}^2 + f_x^2+f_y^2}+\sqrt{f_x^2+f_y^2})} \hat{z}\right) + \sin k_x \frac{f_{Ag} f_x}{f_{Ag}^2+f_x^2+f_y^2+\sqrt{(f_x^2+f_y^2)(f_{Ag}^2+f_x^2+f_y^2)}}$\\
    $ \tau_z  (i\sigma_y)$ & $B_{1u}$& $\frac{f_x^2+f_y^2+\sqrt{(f_x^2+f_y^2)(f_{Ag}^2+f_x^2+f_y^2)}}{f_{Ag}^2+f_x^2+f_y^2+\sqrt{(f_x^2+f_y^2)(f_{Ag}^2+f_x^2+f_y^2)}}$\\
    $ \tau_0 k_z \sigma_y (i\sigma_y)$ & $B_{3u}$ & $\sin k_z\left(\frac{f_x}{\sqrt{f_x^2+f_y^2}}\hat{z} + \frac{f_{Ag}^2 f_y}{\sqrt{f_x^2+f_y^2}(f_{Ag}^2+f_x^2+f_y^2+\sqrt{(f_x^2+f_y^2)(f_{Ag}^2+f_x^2+f_y^2)})} \hat{x}\right)$  \\
   $ \tau_0 k_y \sigma_z (i\sigma_y)$ & $B_{3u}$  & $\sin k_y\frac{f_{Ag}}{\sqrt{f_{Ag}^2+f_x^2 +f_y^2}}\hat{x}$\\
    $ \tau_x k_z \sigma_y (i\sigma_y)$ & $B_{3u}$ & $\sin k_z\left(-\frac{f_y}{\sqrt{f_x^2+f_y^2}}\hat{y} + \frac{f_{Ag}f_x}{\sqrt{f_x^2+f_y^2}(\sqrt{f_{Ag}^2 + f_x^2+f_y^2}+\sqrt{f_x^2+f_y^2})} \hat{z}\right) + \sin k_z \frac{f_{Ag} f_x}{f_{Ag}^2+f_x^2+f_y^2+\sqrt{(f_x^2+f_y^2)(f_{Ag}^2+f_x^2+f_y^2)}}$\\
   $ \tau_x k_y \sigma_z (i\sigma_y)$ & $B_{3u}$  & $\sin k_y \hat{x}$\\
   $ \tau_y \sigma_y (i\sigma_y)$ & $B_{3u}$ & $\frac{f_y}{\sqrt{(f_{Ag}^2+f_x^2 +f_y^2)}} \hat{x} \sim \sin k_y \hat{x}$
    \end{tabular}
    \caption{Momentum dependence of the low-field phase: $B_u$ (in-field) basis functions in the band basis.}
    \label{tab:nodes_Bu}
\end{table}
\begin{table}
    \centering
    \begin{tabular}{c|cc}
   $\Delta$ (SO basis) & $D_{2h}$ Classification & Momentum dependence \\
    \hline
   $ \tau_0 k_x \sigma_x (i\sigma_y)$ & $A_u$ & $\sin k_x\left( \frac{f_y}{\sqrt{f_x^2+f_y^2}}\hat{z} + \frac{f_{Ag}^2 f_x}{\sqrt{f_x^2+f_y^2}(f_{Ag}^2+f_x^2+f_y^2+\sqrt{(f_x^2+f_y^2)(f_{Ag}^2+f_x^2+f_y^2)}}\hat{y}\right)$\\
    $ \tau_0 k_y \sigma_y (i\sigma_y)$ & $A_u$ & $\sin k_y\left(\frac{f_x}{\sqrt{f_x^2+f_y^2}}\hat{z} + \frac{f_{Ag}^2 f_y}{\sqrt{f_x^2+f_y^2}(f_{Ag}^2+f_x^2+f_y^2+\sqrt{(f_x^2+f_y^2)(f_{Ag}^2+f_x^2+f_y^2)})} \hat{x}\right)$ \\
   $ \tau_0 k_z \sigma_z (i\sigma_y)$ &  $A_u$ & $\sin k_z \frac{f_{Ag}}{\sqrt{f_{Ag}^2+f_x^2 +f_y^2}}\hat{x}$\\
   $ \tau_x k_x \sigma_x (i\sigma_y)$ & $A_u$ & $\sin k_x\left(\frac{f_x}{\sqrt{f_x^2+f_y^2}}\hat{y} + \frac{f_{Ag}f_y}{\sqrt{f_x^2+f_y^2}(\sqrt{f_{Ag}^2 + f_x^2+f_y^2}+\sqrt{f_x^2+f_y^2})} \hat{z}\right) + \sin k_x \frac{f_{Ag} f_y}{f_{Ag}^2+f_x^2+f_y^2+\sqrt{(f_x^2+f_y^2)(f_{Ag}^2+f_x^2+f_y^2)}}$\\
    $ \tau_x k_y \sigma_y (i\sigma_y)$ & $A_u$ & $\sin k_y\left(-\frac{f_y}{\sqrt{f_x^2+f_y^2}}\hat{y} + \frac{f_{Ag}f_x}{\sqrt{f_x^2+f_y^2}(\sqrt{f_{Ag}^2 + f_x^2+f_y^2}+\sqrt{f_x^2+f_y^2})} \hat{z}\right) + \sin k_y \frac{f_{Ag} f_x}{f_{Ag}^2+f_x^2+f_y^2+\sqrt{(f_x^2+f_y^2)(f_{Ag}^2+f_x^2+f_y^2)}}$\\
   $ \tau_x k_z \sigma_z (i\sigma_y)$ &  $A_u$ & $\sin k_z \hat{x}$\\
   $ \tau_y \sigma_z (i\sigma_y)$ &  $A_u$ &  $\frac{\sqrt{f_x^2+f_y^2}}{\sqrt{(f_{Ag}^2+f_x^2 +f_y^2)}} \hat{x}$   \\
$ \tau_0 k_z \sigma_x (i\sigma_y)$ & $B_{2u}$ & $\sin k_z\left( \frac{f_y}{\sqrt{f_x^2+f_y^2}}\hat{z} + \frac{f_{Ag}^2 f_x}{\sqrt{f_x^2+f_y^2}(f_{Ag}^2+f_x^2+f_y^2+\sqrt{(f_x^2+f_y^2)(f_{Ag}^2+f_x^2+f_y^2)}}\hat{y}\right)$\\
$ \tau_0 k_x \sigma_z (i\sigma_y)$ &  $B_{2u}$ & $\sin k_x \frac{f_{Ag}}{\sqrt{f_{Ag}^2+f_x^2 +f_y^2}}\hat{x}$\\
  $ \tau_x k_z \sigma_x (i\sigma_y)$ & $B_{2u}$ & $\sin k_z\left(\frac{f_x}{\sqrt{f_x^2+f_y^2}}\hat{y} + \frac{f_{Ag}f_y}{\sqrt{f_x^2+f_y^2}(\sqrt{f_{Ag}^2 + f_x^2+f_y^2}+\sqrt{f_x^2+f_y^2})} \hat{z}\right) + \sin k_z \frac{f_{Ag} f_y}{f_{Ag}^2+f_x^2+f_y^2+\sqrt{(f_x^2+f_y^2)(f_{Ag}^2+f_x^2+f_y^2)}}$\\
 $ \tau_x k_x \sigma_z (i\sigma_y)$ &  $B_{2u}$ & $\sin k_x \hat{x}$\\
$ \tau_y \sigma_x (i\sigma_y)$ & $B_{2u}$&  $\frac{f_x}{\sqrt{(f_{Ag}^2+f_x^2 +f_y^2)}} \hat{x} \sim \sin k_x \hat{x}$\\
    \end{tabular}
    \caption{Momentum dependence of the high-field phase: $A_u$ (in-field) basis functions in the band basis}
    \label{tab:nodes_Au}
\end{table}

\end{widetext}

\end{document}